\documentclass[preprint,aps,showpacs]{revtex4}

\begin{document}
\title{The derivation of Bell inequalities for beables}
\author{Peter Morgan}
         \affiliation{30, Shelley Road, Oxford, OX4 3EB, England.}
         \email{peter.morgan@philosophy.oxford.ac.uk}
         \altaffiliation{Address until August 2003:
                             207, von Neumann Drive, Princeton NJ08540, USA.}
         \homepage{http://users.ox.ac.uk/~sfop0045}
\date{\today}

\begin{abstract}
The derivation of Bell inequalities for beables is well-known to require a
``no-conspiracy'' assumption.
This assumption is widely accepted, the alternative being correlations between
instrument settings and hidden beables.
Two further assumptions are identified here: 
(1) a ``no-contextuality'' assumption, similar to the prohibition of contextuality that is
required to derive the Kochen-Specker theorem, which is closely related to the
``no-conspiracy'' assumption;
(2) a ``no-correlation'' assumption, which prohibits correlations between hidden
beables.
The three assumptions together are less acceptable than the
``no-conspiracy'' assumption alone.
\end{abstract}

\newcommand{\RA}{\mathcal{R}_A}
\newcommand{\RB}{\mathcal{R}_B}
\newcommand{\PastRA}{\textsl{Past}(\RA)}
\newcommand{\PastRB}{\textsl{Past}(\RB)}
\newcommand{\Note}[1]{{\textbf{[[}\textsl{#1}\textbf{]]}}}
\maketitle

\setlength{\parskip}{4pt}
\section{Introduction}
I take it that the violation of Bell inequalities by experiment rules out any simple local
beables model for the violation of Bell inequalities that is constructed using the resources
of \emph{classical particle mechanics} (although conceptually  extravagant notions of
backward causal propagation, for example, are not ruled out, and detector
efficiency remains an issue), but I consider here the violation of Bell
inequalities in local beables models that are constructed using the resources
of \emph{classical statistical field theory}. Nonlocal correlations are always present
in the context of classical statistical field theory, whereas they are generally not
present in the classical mechanics of a small number of particles.

There are other reasons for thinking that beables models must be nonlocal,
particularly Hegerfeldt nonlocality\cite{Hegerfeldt, Morgan}, but Bell inequalities
for beables do not provide a strong constraint on classical statistical fields as a
basis for a beables model for the violation of Bell inequalities.

Bell\cite[Chap. 7, originally 1976]{Bell} shows from a definition of local causality alone that a
beables model predicts two Bell inequalities for observable classical statistics
associated with two space-like separated regions $\RA$ and $\RB$, and that
quantum theory does not satisfy the same inequalities.
\begin{figure}[htb]
\setlength{\unitlength}{0.038\columnwidth}
\begin{picture}(26,9) 
\multiput(0,2.4)(9,0){2}{\line(1,1){6}}
\multiput(10,8.4)(9,0){2}{\line(1,-1){6}}
\multiput(0,2.4)(9,0){2}{\qbezier(0,0)(0,1)(8,1)}
\multiput(0,2.4)(9,0){2}{\qbezier(8,1)(16,1)(16,0)}
\multiput(0,2.4)(9,0){2}{\qbezier(0,0)(0,-1)(8,-1)}
\multiput(0,2.4)(9,0){2}{\qbezier(8,-1)(16,-1)(16,0)}
\multiput(6,8.4)(9,0){2}{\qbezier(0,0)(0,.5)(2,.5)}
\multiput(6,8.4)(9,0){2}{\qbezier(2,.5)(4,.5)(4,0)}
\multiput(6,8.4)(9,0){2}{\qbezier(0,0)(0,-.5)(2,-.5)}
\multiput(6,8.4)(9,0){2}{\qbezier(2,-.5)(4,-.5)(4,0)}
\put(7.5,8.15){$\RA$}
\put(16.5,8.15){$\RB$}
\put(7,5.4){$\PastRA$}
\put(16,5.4){$\PastRB$}
\put(16,0){\framebox{$\PastRA\cap\PastRB$}}
\put(16,.3){\vector(-3,4){3.2}}
\put(5,2.2){$a,\lambda$}
\put(18.5,2.2){$b,\mu$}
\put(11.7,2.2){$c,\nu$}
\end{picture}
\caption{\label{Fig1}Space-time regions and their beables.}
\end{figure}
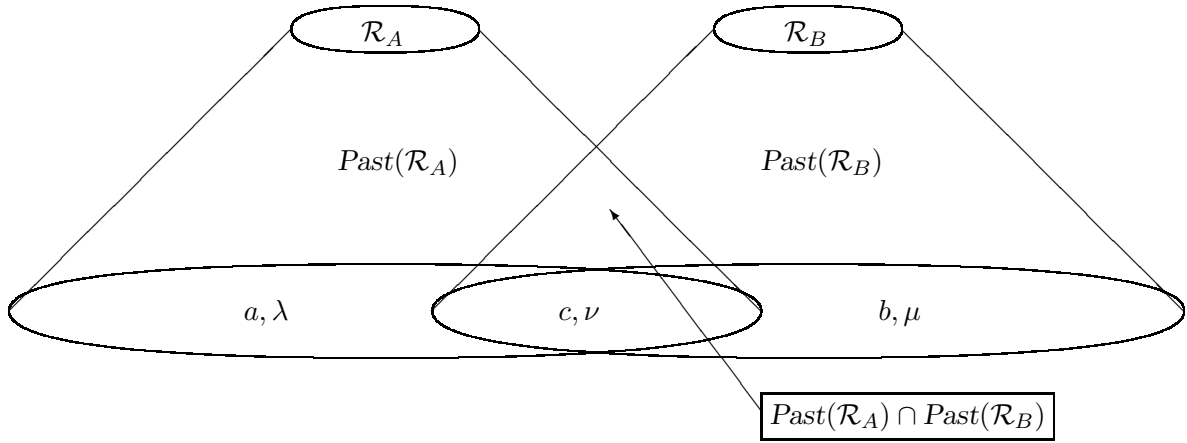
Shimony, Horne, and Clauser\cite[originally 1976]{SHC} show, however, that  if
beables associated with $\PastRA - \PastRB$ and with $\PastRB - \PastRA$
are correlated with hidden beables associated with $\PastRA \cap \PastRB$,
which we cannot rule out on causal grounds because the forward light cone of
$\PastRA \cap \PastRB$ fills all of space-time, then the beables model need not
satisfy the Bell inequalities.
Bell\cite[Chap. 12, originally 1977]{Bell} admits this, but finds that hidden beables associated
with $\PastRA \cap \PastRB$ have to be correlated with instrument settings in $\RA$
and in $\RB$.
Arguing that such a requirement is unreasonable, Bell calls it a ``conspiracy''.
Bell's argument and Shimony, Horne, and Clauser's comments are brought
together in a review article by d'Espagnat\cite{dEspagnat}, and Brans\cite{Brans}
gives an alternative, quite helpful discussion.
The literature on Bell inequalities \emph{for beables} is quite sparse (very sparse indeed by
comparison with the literature on Bell inequalities more generally), and has not changed
the general perception that Bell\cite[Chap. 12]{Bell} more-or-less closes the discussion.

The beables $(a,\lambda)$, $(b,\mu)$, and $(c,\nu)$ are associated with the
disjoint regions $\PastRA - \PastRB$, $\PastRB - \PastRA$, and
$\PastRA\cap \PastRB$, respectively (see figure \ref{Fig1}).
$a$, $b$, and $c$ are ``non-hidden''\cite[Chap. 12]{Bell} beables, instrument settings
that are observed and possibly controlled by the experimenter, while $\lambda$, $\mu$,
and $\nu$ are hidden beables, neither observed nor controlled by the experimenter.
As far as classical physics is concerned, the separation of beables into $(a,\lambda)$,
$(b,\mu)$, and $(c,\nu)$ is arbitrary, because anything that is hidden today may be
revealed tomorrow and whether we observe or record beables makes no difference,
so any derivation of Bell inequalities must be robust under different choices of the
separation.
There is nothing about the mathematics of section \ref{Derivation} that will determine
a separation of beables into $(a,\lambda)$, $(b,\mu)$, and $(c,\nu)$.
The only difference between non-hidden beables and hidden beables will be that we
will integrate over all values of hidden beables and never integrate over non-hidden
beables.
It will be useful to consider three choices in this paper: (1) all of $a$, $b$, $c$,
$\lambda$, $\mu$, and $\nu$ non-null; (2) $\nu$ maximal, so that $c$ is null; and
(3) $c$ maximal, so that $\nu$ is null.

The fundamental definition in Bell's derivation of inequalities for beables is that
for a \emph{locally causal theory}, for $X$ any beables associated with
$\RA$, $X_\cap$ \emph{all} of the beables associated with
$\PastRA \cap \PastRB$, $X_p$ \emph{some} of the beables
associated with $\PastRA - \PastRB$, and $Y$ any beables
associated with $\RB$, the conditional probability of $X$ given $X_\cap$
and $X_p$ is statistically independent of $Y$,
\begin{equation}\label{LocalCausality}
  p(X|X_\cap, X_p, Y)=p(X|X_\cap, X_p).
\end{equation}
(In an alternative terminology, usual in philosophy, correlation between
$X$ and $Y$ is ``screened off'' by $X_\cap$ and $X_p$.)
This definition is applied a number of times in Bell's derivation of inequalities
for beables.
Shimony, Horne, and Clauser\cite{SHC}, in contrast, weaken the definition of a locally
causal theory, so that for $X$ and $Y$ as above, but for $X_P$ \emph{all} of the
beables associated with $\PastRA$, the conditional probability of $X$
given $X_P$ is statistically independent of $Y$,
\begin{equation}\label{LocalCausalitySHC}
  p(X|X_P, Y)=p(X|X_P).
\end{equation}
(Correlation between $X$ and $Y$ is ``screened off'' by $X_P$.)
The two definitions are the same if $X_p$ happens to be a specification of \emph{all}
the beables in $\PastRA - \PastRB$.
Equation (\ref{LocalCausality}) combines equation (\ref{LocalCausalitySHC}), which
is a more natural definition of local causality, with a principle that correlations only
arise because of common causes.
Equation (\ref{LocalCausalitySHC}), however, is not strong enough to allow Bell
inequalities to be derived.
Some of the applications of equation (\ref{LocalCausality}) can be replaced by
applications of equation (\ref{LocalCausalitySHC}), but most cannot.
One of the latter applications is the ``no-conspiracy'' assumption, which prohibits
correlations between instrument settings and hidden beables, and section
\ref{Derivation} further identifies a ``no-correlation'' assumption (discussed in section
\ref{CorrelationSection}), which prohibits correlations between hidden beables.

The derivation of Bell inequalities for beables also requires a previously unidentified
assumption that a beables theory is noncontextual. 
Section \ref{Derivation} identifies a ``no-contextuality'' assumption (discussed in section
\ref{ConspiracySection}), similar to the prohibition of contextuality that is required of a
beables model to derive the Kochen-Specker theorem\cite{Redhead}, and shows the
``no-contextuality'' assumption to be closely related to the ``no-conspiracy'' assumption.
The three assumptions taken together are much less acceptable than the
``no-conspiracy'' assumption alone.

Finally, section \ref{QFTApproach} shows that the violation of Bell inequalities alone
does not justify preferring an empiricist interpretation of quantum field theory over
an equally empiricist interpretation of classical statistical field theory by considering
the similarities between a quantum field theoretic Wigner quasi-probability description
and a classical probability description of an experiment that violates a Bell inequality.

\section{Derivation of Bell inequalities for beables}
\label{Derivation}
Bell's mathematical argument is reproduced here in the form given by
d'Espagnat\cite{dEspagnat}.
\Note{Notes in brackets} will indicate where there are assumptions that will be
addressed in the next two sections.
Suppose that $A$ and $B$ are observed in regions $\RA$ and $\RB$.
Applying equation (\ref{LocalCausality}) \emph{or} equation (\ref{LocalCausalitySHC}),
the conditional probability density $p(A|a,b,c,\lambda,\mu,\nu,B)$ is
statistically independent of $b$, $\mu$, and $B$ in a locally causal theory,
and similarly for the conditional probability density
$p(B|a,b,c,\lambda,\mu,\nu,A)$,
\begin{eqnarray}
   p(A|a,b,c,\lambda,\mu,\nu,B) &=& p(A|a,c,\lambda,\nu),\\
   p(B|a,b,c,\lambda,\mu,\nu,A) &=& p(B|b,c,\mu,\nu).
\end{eqnarray}
The mean of the product $AB$, given the non-hidden beables $(a,b,c)$, is:
\begin{eqnarray}
M(a,b,c) & = & \int\!\!\!\int\!\!\!\int\sum_{AB}
                    AB\,p(A,B,\lambda,\mu,\nu|a,b,c) d\lambda d\mu d\nu \cr
         & = & \int\!\!\!\int\!\!\!\int\sum_{AB}
                    AB\,p(A|a,c,\lambda,\nu)p(B|b,c,\mu,\nu)
                    p(\lambda,\mu,\nu|a,b,c) d\lambda d\mu d\nu,
\end{eqnarray}
where the conditional probability density $p(\lambda,\mu,\nu|a,b,c)$ for the hidden
beables can be rewritten, applying no more than the definition of conditional
probability, as
\begin{eqnarray}
     p(\lambda,\mu,\nu|a,b,c) &\! = \!& p(\lambda|\mu,\nu,a,b,c)p(\mu,\nu|a,b,c)\cr
     {} &\! = \!& p(\lambda|\mu,\nu,a,b,c)p(\mu|\nu,a,b,c)p(\nu|a,b,c).
\end{eqnarray}
Applying equation (\ref{LocalCausality}), or, through a putative argument provided by
Shimony, Horne, and Clauser\cite{SHC} and discussed in section \ref{CorrelationSection},
applying equation (\ref{LocalCausalitySHC}), we can derive
\begin{eqnarray}
    p(\lambda|\mu,\nu,a,b,c) &=& p(\lambda|\nu,a,b,c), \label{No-Correlation}\\
    p(\lambda|\nu,a,b,c) &=& p(\lambda|\nu,a,c), \label{No-NonlocalA}\\
    p(\mu|\nu,a,b,c) &=& p(\mu|\nu,b,c) \label{No-NonlocalB}
\end{eqnarray}
\Note{$p(\lambda|\mu,\nu,a,b,c) = p(\lambda|\nu,a,b,c)$ is the ``no-correlation''
     assumption; equations (\ref{No-NonlocalA}) and (\ref{No-NonlocalB}) are
     further assumptions, which might be called ``no-nonlocal-conspiracy''
     assumptions, but will not be directly addressed here},
so the mean of the product $AB$, given non-hidden beables $(a,b,c)$, is
\begin{equation}\label{MeanProduct}
    M(a,b,c) = \int\overline{A(a,c,\nu)}\>\overline{B(b,c,\nu)}p(\nu|a,b,c)d\nu,
\end{equation}
where $\overline{A(a,c,\nu)}$ is the mean of $A$ averaged over the
hidden beables $\lambda$, given the non-hidden beables $a$ and $c$ and
the hidden beables $\nu$, and similarly for $\overline{B(b,c,\nu)}$.

Suppose that $A$ and $B$ satisfy $|A|\leq 1$ and $|B|\leq 1$, so that
$|\overline{A(a,c,\nu)}|\leq 1$ and $|\overline{B(b,c,\nu)}|\leq 1$. If we
also suppose that
\begin{equation}\label{ConspiracyEquation}
   p(\nu|a,b,c) = p(\nu|c)
\end{equation}
\Note{$p(\nu|a,b,c) = p(\nu|c)$ is the already known ``no-conspiracy'' assumption},
then we can derive, for distinct non-hidden beables $a$, $a'$, and $b$, $b'$,
\begin{eqnarray}
    |M(a,b,c)\mp M(a,b',c)| & = & \left|\int\overline{A(a,c,\nu)}
      \left[\overline{B(b,c,\nu)}\mp \overline{B(b',c,\nu)}\right]
                  p(\nu|c)d\nu\right| \cr
 \noalign{\vskip 3pt}
    {} & \leq &
    \left|\int\left[\overline{B(b,c,\nu)}\mp \overline{B(b',c,\nu)}\right]
                  p(\nu|c)d\nu\right|,\\
 \noalign{\vskip 6pt}
    |M(a',b,c)\pm M(a',b',c)| & = & \left|\int\overline{A(a',c,\nu)}
      \left[\overline{B(b,c,\nu)}\pm \overline{B(b',c,\nu)}\right]
                  p(\nu|c)d\nu\right| \cr
 \noalign{\vskip 3pt}
    {} & \leq &
    \left|\int\left[\overline{B(b,c,\nu)}\pm \overline{B(b',c,\nu)}\right]
                  p(\nu|c)d\nu\right|,
\end{eqnarray}
\Note{Being able to change $a\rightarrow a'$ without changing $c$ or $b$ and
$b\rightarrow b'$ without changing $c$ or $a$ is the ``no-contextuality'' assumption,
so-called here because it signifies that state preparation and measurement devices
are independent} so that
\begin{equation}\label{BellInequality}
    |M(a,b,c)\mp M(a,b',c)|+|M(a',b,c)\pm M(a',b',c)|\leq 2.
\end{equation}
In contrast, for two spin-half particles, we can derive the inequalities
\begin{equation}
    |M(a,b,c)\mp M(a,b',c)|+|M(a',b,c)\pm M(a',b',c)|\leq 2\sqrt{2},
\end{equation}
\begin{tabular}{c l l} 
{but,} & {if $p(\lambda|\mu,\nu,a,b,c) \ne p(\lambda|\nu,a,b,c)$,} & {--- correlation}\\
{} &       {or $p(\lambda|\nu,a,b,c) \ne p(\lambda|\nu,a,c)$,} & {--- nonlocal-conspiracy}\\
{} &       {or $p(\mu|\nu,a,b,c) \ne p(\mu|\nu,b,c)$,} & {--- nonlocal-conspiracy}\\
{} &       {or $p(\nu|a,b,c)\ne p(\nu|c)$,} & {--- conspiracy}\\
\noalign{\vskip -1pt}
{} &       {or if we cannot change $a$, $b$, and $c$ independently,\hspace{2em}} & {--- contextuality}\\
\noalign{\vskip 5pt}
\end{tabular}\newline
then we can only derive the inequalities
\begin{equation}\label{MaxInequalities}
  |M(a,b,c)\mp M(a,b',c)|+|M(a',b,c)\pm M(a',b',c)|\leq 4.
\end{equation}
Classically, non-relativistic quantum mechanics is half-way between the
conditions for deriving Bell inequalities and the maximum violation, when
equation (\ref{MaxInequalities}) is satisfied as an equality.
There must, therefore, be principled constraints on initial conditions in a beables
model to ensure the maximum violation is never observed, as well as to allow
some violation.

\section{The no-conspiracy and no-contextuality assumptions}
\label{ConspiracySection}
The prohibition of correlations of $a$ with $c$, and of $b$ with $c$,
the ``no-contextuality'' assumption, is closely related to the ``no-conspiracy''
assumption.
If we suppose that $\nu$ is complete information, so that $c$ is null, we can
derive in place of equation (\ref{MeanProduct}), supposing that equations
(\ref{No-Correlation}), (\ref{No-NonlocalA}), and (\ref{No-NonlocalB}) are satisfied,
\begin{equation}
    M(a,b) = \int\overline{A(a,\nu)}\>\overline{B(b,\nu)}p(\nu|a,b)d\nu,
\end{equation}
which requires that
\begin{equation}
    p(\nu|a,b)=p(\nu)
\end{equation}
for us to be able to derive Bell inequalities.
If we take $a$ and $b$ to be only instrument settings at the time of the
measurement, with $c$ null, the ``no-conspiracy'' assumption is little different
than to say that
\emph{instrument settings at the time of the measurement must be} completely
\emph{uncorrelated with the experimental apparatus} (which is, after all, almost
entirely in $\PastRA\cap \PastRB$).
Ensuring that instrument settings are completely uncorrelated with the experimental
apparatus would seem a remarkable achievement in a classical statistical field theory
setting.

Bell argues\cite[Chap. 12]{Bell} that the dynamics of a mechanism to choose the
instrument settings can be made chaotic enough that, even if there are correlations
between $(c,\nu)$ and $(a, b)$, the instrument settings may nonetheless be taken
to be ``at least effectively free for the purposes at hand''.
From a classical point of view, this is a remarkable claim.
Either there are correlations in a model for an experiment or there are not.
Correlations that are easy to measure at one time are generally not as easy to measure
at other times, but the practicality of measuring correlations has no bearing on whether
there are correlations, which is in principle unaffected by whether the evolution is
chaotic or not.

In any case, $a$ and $b$ being ``free for the purposes at hand'' does \emph{not}
imply $p(\nu|a,b)=p(\nu)$. A correlation $p(\nu|a,b)\not=p(\nu)$ does not
``determine'' $a$ and $b$ (or $\nu$), but only describes a statistical relationship
between $a$, $b$, and $\nu$.
 
Bell also argues\cite[Chap. 12]{Bell}
``that the disagreement between locality and quantum mechanics is large --- up to
a factor of $\sqrt{2}$ in a certain sense''. The standard assumptions discussed here
are given as analytic equalities, which are unable to elaborate the ``certain sense''.
Beables are so general, however, that it is unclear how no-correlation,
no-nonlocal-conspiracy, no-conspiracy, and no-contextuality could be given as physically
justifiable limits on inequality (note that the standard assumptions are problematic just
as analytic equalities between probability distributions, since such a relationship cannot
be supported by experimental statistics, nor, it seems, by analytic argument).
Bell's argument is also weakened by the classical limit being either $2$ or $4$ (Bell
omits to mention the latter), depending on whether we accept \emph{all} the standard
assumptions, with $2\sqrt{2}$ as the intermediate quantum mechanical limit.

Suppose that instead of taking $\nu$ to be complete information, we take $c$ to be
complete information, so that $\nu$ is null.
Then we can derive, in place of equation
(\ref{MeanProduct}), again supposing that equations (\ref{No-Correlation}),
(\ref{No-NonlocalA}), and (\ref{No-NonlocalB}) are satisfied,
\begin{equation}
M(a,b,c)=\overline{A(a,c)}\ \overline{B(b,c)}.
\end{equation}
Now to derive equation (\ref{BellInequality}), we have to be able to change
$a\rightarrow a'$ without changing $c$ (or $b$) and be able to change
$b\rightarrow b'$ without changing $c$ (or $a$), with the ``no-conspiracy''
assumption playing no r\^ole.
Nonlocality aside, requiring this is the ``no-contextuality'' assumption.
Note that it has been argued before that Bell inequalities have little or nothing
to do with nonlocality\cite{deMuynck}, but are instead a consequence only of an
assumption that quadrivariate probability distributions can be introduced for
incompatible observables, which can be avoided only by introducing contextual
classical models.

In quantum field theory, the Reeh-Schlieder theorem\cite{Haag} is typically
thought very awkward, yet the contextuality it implies is not taken to rule out
quantum field theory.
Recall that as a consequence of the Reeh-Schlieder theorem we cannot change
a quantum field state so that the expected value of a quantum field observable
associated with $\PastRA - \PastRB$ changes without changing  the expected
value of at least some quantum field observables associated with both
$\PastRA\cap\PastRB$ and $\PastRB - \PastRA$.
In classical statistical terms this is just to say that we cannot generally change
$a\rightarrow a'$ without changing $b$ and $c$ at least some of the time.
If $c$ is complete information about observables in $\PastRA\cap\PastRB$, this
would be impossible in quantum field theory.

It is unreasonable to expect noncontextuality of a beables model when we do not
expect it of quantum field theory --- to do so is to construct a straw man of a theory.
If we insist on a parallel of the Reeh-Schlieder theorem in classical
statistical field theory, we cannot derive Bell inequalities for beables.

\section{The no-correlation assumption}
\label{CorrelationSection}
Recall that the ``no-correlation'' assumption, equation (\ref{No-Correlation}),
requires that there are no correlations between the hidden beable $\lambda$ and
the hidden beables $\mu$ (that are not screened off by $\nu$, $a$, $b$, and $c$).
There is no empirical way to justify this assumption, simply because it is a condition
imposed on observables that are \emph{by definition} not measured.
The preference against correlations between instrument settings and hidden beables
is only tendentiously extensible to justify a prohibition against correlations between
hidden beables.

Shimony, Horne, and Clauser\cite{SHC} argue that
\begin{quote}
``even though the space-time region in which $\lambda$ is located extends to negative
infinity in time, $\nu$, $a$, $c$ are \emph{all} the beables other than $\lambda$
itself in the backward light cone of this region, and $\mu$ and $b$ \emph{do} refer
to beables with space-like separation from the $\lambda$ region''
\end{quote}
to justify deriving equations (\ref{No-Correlation}), (\ref{No-NonlocalA}), and
(\ref{No-NonlocalB}) from equation (\ref{LocalCausalitySHC}).
This argument is \emph{precisely} to revert to Bell's definition of local
causality, equation (\ref{LocalCausality}) 
[with $X_\cap=c,\nu$; $X_p=a$; $X_P=c,\nu,a,\lambda$].
To try to justify equation (\ref{No-Correlation}), using equation
(\ref{LocalCausalitySHC}), we could write
\begin{equation}\label{TrivialCorrelation}
       p(\lambda|c,\nu,a,\lambda,b,\mu)  =  p(\lambda|c,\nu,a,\lambda,b),
\end{equation}
but it would be rather remarkable to depend on this equation, since
$p(\lambda|\lambda,Z)$ is trivially equal to $p(\lambda|\lambda)$
whatever $Z$ represents, independently of any idea of local causality.
From the trivial equality (\ref{TrivialCorrelation}) we cannot derive
\begin{equation}
       p(\lambda|c,\nu,a,\hspace{1 em}b,\mu)  =  p(\lambda|c,\nu,a,\hspace{1 em}b),
\end{equation}
which is a consequence of equation (\ref{LocalCausality}) but not of equation
(\ref{LocalCausalitySHC}).
If we take the trouble to distinguish between equation (\ref{LocalCausality}) and
equation (\ref{LocalCausalitySHC}) as definitions of local causality, we can hardly
argue for equations (\ref{No-Correlation}), (\ref{No-NonlocalA}), and
(\ref{No-NonlocalB}) like this.

Shimony, Horne, and Clauser's argument depends weakly on $\lambda$ being
associated with the whole of $\PastRA-\PastRB$, so that $\lambda$, $\nu$, $a$,
and $c$ are all the beables in the backward light-cone of $\PastRA-\PastRB$.
Let us consider, therefore, the \emph{slightly} less general class of beables
models that satisfy equation (\ref{LocalCausalitySHC}) and for which also a
probability distribution $p(a_t,b_t,c_t,\lambda_t,\mu_t,\nu_t)$ over non-hidden
and hidden beables associated with a cross-section of $\PastRA$ and
$\PastRB$ at a time $t$ determines the equivalent probability distribution
at later times.
Such a beables model includes both deterministic and indeterministic models,
and both reversible and irreversible dynamics, but does not generally satisfy
equation (\ref{LocalCausality}).
If we suppose that the cross-section is a space-time region of finite duration
$\Delta t$, we can allow slightly more dynamical generality than would be allowed
if we supposed the cross-section to be a hyperplane.
Such a class of beables models includes almost every model we would usually
think of as classical. Bell's construction of models in which $(a,\lambda)$ is a complete
set of beables associated with \emph{all} of $\PastRA$ is something of a novelty.

\begin{figure}[htb]
\setlength{\unitlength}{0.036\columnwidth}
\begin{picture}(26,7)\put(0,-2){
\multiput(0,2.4)(9,0){2}{\line(1,1){6}}
\multiput(10,8.4)(9,0){2}{\line(1,-1){6}}
\multiput(0,2.4)(9,0){2}{\qbezier(0,0)(0,1)(8,1)}
\multiput(0,2.4)(9,0){2}{\qbezier(8,1)(16,1)(16,0)}
\multiput(0,2.4)(9,0){2}{\qbezier(0,0)(0,-1)(8,-1)}
\multiput(0,2.4)(9,0){2}{\qbezier(8,-1)(16,-1)(16,0)}
\multiput(6,8.4)(9,0){2}{\qbezier(0,0)(0,.6)(2,.6)}
\multiput(6,8.4)(9,0){2}{\qbezier(2,.6)(4,.6)(4,0)}
\multiput(6,8.4)(9,0){2}{\qbezier(0,0)(0,-.6)(2,-.6)}
\multiput(6,8.4)(9,0){2}{\qbezier(2,-.6)(4,-.6)(4,0)}
\put(7.2,8.15){$A,s_A$}
\put(16.2,8.15){$B,s_B$}
\put(5,2.2){$a_t,\lambda_t$}
\put(18.5,2.2){$b_t,\mu_t$}
\put(11.7,2.2){$c_t,\nu_t$}
\put(25.5,8.4){$t_M$}
\put(24.7,5.4){$t_M-\tau_\cap$}
\put(25.5,2.2){$t$}
}\end{picture}
\caption{\label{Fig2}Beables associated with hyperplane slices at $t_M$ and $t$.}
\end{figure}
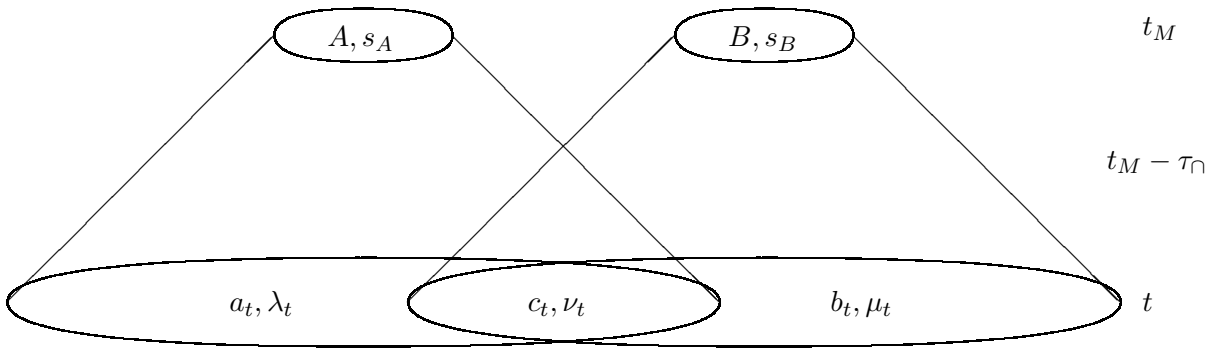
At the time $t_M$ of the measurements in $\RA$ and $\RB$, and for a time
$\tau_\cap$ before the measurement, there are no non-hidden beables $c_t$
or hidden beables $\nu_t$.
Suppose that actually observed experimental statistics over $(A,s_A,B,s_B)$ that
violate Bell inequalities are adequately modelled by a joint probability distribution
$p(A,s_A,B,s_B)$, where $s_A$ and $s_B$ are instrument settings at time $t_M$.
\emph{Any} probability distribution that has $p(A,s_A,B,s_B)$ as a marginal
distribution is an adequate contextual beables model at time $t_M$.
A probability distribution $p(A,s_A,B,s_B)$ is \emph{not} a quadrivariate probability
distribution over incompatible observables, as a probability distribution over
$(A_1,A_2,B_1,B_2)$ would be if it were constructed by post-selecting values of $A$
corresponding to different values of $s_A$ and values of $B$ corresponding to different
values of $s_B$, which would lead directly to Bell-type inequalities\cite{deMuynck}
without any locality assumptions being necessary.
$A$, $s_A$, $B$, and $s_B$ are four mutually compatible observables that have to
take specific values in an experiment (or in a number of experiments) for us to be
able to construct statistics corresponding to $M(a,b,c)$, $M(a,b',c)$, $M(a',b,c)$,
and $M(a',b',c)$ and hence to verify that Bell inequalities are violated by experiment.

From the starting point of $p(A|\lambda_{t_M},a_{t_M})$ and  $p(B|\mu_{t_M},b_{t_M})$,
if we assume ``no-correlation'' and ``no-nonlocal-conspiracy'',
\begin{eqnarray}
  \left.\begin{array}{r c l}
    p(\lambda_{t_M}|\mu_{t_M},s_A,s_B)&\!\!=\!\!&p(\lambda_{t_M}|s_A,s_B)\\
    p(\lambda_{t_M}|s_A,s_B)&\!\!=\!\!&p(\lambda_{t_M}|s_A)\\
    p(\mu_{t_M}|s_A,s_B)&\!\!=\!\!&p(\mu_{t_M}|s_B)
  \end{array}\!\!\right\}\longrightarrow
  M(s_A,s_B)=\overline{A(s_A)}\ \overline{B(s_B)}.\hspace{-1em}\cr
  \noalign{\vskip -1em}
\end{eqnarray}
Just ``no-contextuality'', without ``no-conspiracy'', proves that
Bell inequalities must be satisfied.
It is straightforward, however, to construct a hidden beables model for
$p(A,s_A,B,s_B)$ by just measuring additional observables that are compatible
with $A$,$s_A$,$B$, and $s_B$, in an experiment for which Bell inequalities
are not satisfied.
For example, we could measure details of the thermodynamical states of whatever
detectors we are using, replacing $A$ by other classical information, from which
$A$ can be derived, or we could replace $s_A$ by measurements of geometrical
details of the apparatus that is used to achieve varied polarizations, from which
$s_A$ can be derived.
Since such a hidden beables model is derived from measurements of an apparatus
in which Bell inequalities are not satisfied, at least some of the assumptions
required to derive Bell inequalities for beables must not be satisfied for such
a model.

At times before $t_M-\tau_\cap$, the set of non-hidden beables $c_t$ and hidden beables
$\nu_t$ is nonempty and increases as $t$ refers to earlier times, and presumably the
assumptions come closer to being satisfied.
At earlier times, however, there is no requirement that the assumptions be violated
by much, only that the totality of correlations be such that the dynamical
evolution will result in the violation of Bell inequalities at the time of
measurement, so the constraint on a beables model is insignificant.

\section{A quantum field theory approach}
\label{QFTApproach}
We have become used to describing the outcome of Bell violating experiments
using a state in a complex 4-dimensional Hilbert space, in which many detailed
degrees of freedom are integrated out.
If we agree, however, that non-relativistic quantum mechanics is a reduction
of quantum field theory, as we almost always do, such a state is a reduction of
a quantum field state in an infinite-dimensional Hilbert space, which gives the
values of quantum field observables associated with the regions $\RA$ and $\RB$.
If Bell inequalities are violated by observables of a quantum field state, we
would certainly attribute the violation to the experimenters' ingenuity in
ensuring an appropriate initial quantum field state and making appropriate
measurements.
For a quantum field state describing an experiment that violates Bell inequalities,
the existence of nontrivial correlations between observables at large
space-like separations is precisely what singles out such states as special.
A quantum field state that describes experimental correlations that violate Bell
inequalities at the time of measurement describes correlations in the remote past
different from those of the vacuum state, but, as for a beables model, differences
from the vacuum state may be difficult to detect in the remote past.
In quasi-probability terms, we have to set up a Wigner quasi-distribution over
phase space in the past that evolves to a Wigner quasi-distribution over phase
space at time $t_M$ that violates a Bell inequality in the regions $\RA$ and $\RB$.

For an equilibrium state of a classical statistical field theory, correlations that
violate the assumption of statistical independence at space-like separation
decrease more-or-less exponentially fast with increasing distance, but non-trivial
correlations at arbitrarily large distances are possible for non-equilibrium states.
Indeed, absolutely any correlations are allowed in a non-equilibrium initial
condition --- initial conditions of low probability of course require greater free
energy to set up, but we should not forget how difficult it is to construct an
experiment that violates Bell inequalities.
In a classical statistical field theory, we have to set up a probability distribution over
phase space in the past that evolves to a probability distribution over phase space at
time $t_M$ that violates a Bell inequality in the regions $\RA$ and $\RB$,
but this is no greater ``conspiracy'' than is apparent in the full quantum field state
for the experiment.
Hence the violation of Bell inequalities does not provide a justification for preferring
quantum theoretical description over classical theoretical description.

Since the measurements required to violate Bell inequalities experimentally require
only that we measure compatible observables $(A,s_A,B,s_B)$ in a single
experimental context, the quantum field state is not fixed uniquely by observation.
In particular, for a single experimental context there is a non-empty set of
quantum field states that have a positive semi-definite Wigner distribution and are
empirically adequate, and we can adopt any of these quantum field states, or rather
any of the associated positive semi-definite Wigner distributions, as a classical model
for the experiment.

The correlations we have discussed here commit us to very little. If we take an
equally empiricist approach to classical statistical field theory as we take to quantum
field theory, there just are correlations, which we don't have to assume are caused by
common (or any other kind of) causes. Some correlations just are.
This does not preclude asking whether we can construct a model that explains the
correlations, but this can reasonably be a future enterprise.
Classical physics has always taken initial conditions to be more-or-less explained by
earlier initial conditions, with no final explanation being essential.
Although of course an explanation of \emph{why} initial conditions are the way they
are may well be more helpful, a description of \emph{what} the initial conditions are
is a helpful enough beginning.

\section{Conclusion}
We have first described the previously identified difference between Bell's definition
of a locally causal theory, which insists that correlations have to be the result of common
causes, and Shimony, Horne, and Clauser's definition, which does not.
Secondly, in a significant change from both Bell's account and Shimony, Horne, and
Clauser's account, we have identified the ``no-contextuality'' assumption, which effectively
extends the ``no-conspiracy'' assumption far enough to make it unacceptable.
Thirdly, in another significant change from both Bell's account and Shimony, Horne, and
Clauser's account, we have identified the ``no-correlation'' assumption, which changes the
type of correlations that are prohibited to include correlations between hidden
beables.

Bell accepts\cite[Chapter 12, last paragraph]{Bell} that
  ``A theory may appear in which such conspiracies inevitably occur, and these
conspiracies may then seem more digestible than the non-localities of other theories'',
but concludes
  ``But I will not myself try to make such a theory''.
For those who \emph{would} construct hidden-variable models, a theory of
such models must include one or all of conspiracies, correlations, and contextuality
in a principled way.
With the extra flexibility offered by denying the several assumptions identified here, it
seems more reasonable to try, at least if we allow ourselves to use the resources of
classical statistical field theory (in contrast to using only the classical mechanics of a
small number of particles, where the correlations needed are not very natural).
Quantum field theory gives us a simple and effective way to proceed, by looking for
all quantum field states that describe our knowledge of a complete experiment, and
accepting only those quantum field states that have a representation as a positive
semi-definite Wigner distribution.

To temper the localism of this article, repeating the caution given in the
introduction, there are reasons for thinking that beables models must be
nonlocal, particularly Hegerfeldt nonlocality\cite{Hegerfeldt, Morgan}, but the
violation of Bell inequalities can be modelled adequately by local beables.

\begin{acknowledgments}
I am grateful to Marcus Appleby, Stephen Adler, Katherine Brading, Harvey Brown,
Jeremy Butterfield, Nancy Cartwright, Willem de Muynck, Sheldon Goldstein,
Jonathan Halliwell, Lucien Hardy, Fred Kronz, James Ladyman, Andrew Laidlaw,
Tracy Lupher, Bert Schroer, John Schutz, Mauricio Suarez, Victor Suchar,
Caroline Thompson, Chris Timpson, Lev Vaidman, David Wallace, Andrew Whitaker,
and to a referee for conversation and criticism.
\end{acknowledgments}

\end{document}